\documentclass[
superscriptaddress,
reprint,
amsmath,amssymb,
aps,
pre,
longbibliography
]{revtex4-1}

\usepackage{graphicx}
\usepackage{epstopdf}
\usepackage{dcolumn}
\usepackage{bm}
\usepackage{xcolor}
\usepackage[colorlinks=true,citecolor=blue,linkcolor=magenta]{hyperref}

\begin{document}


\title{Permutation Phase and Gentile Statistics}

\author{Qiang Zhang}\email{q.zhang@iphy.ac.cn}
\affiliation{%
Beijing National Laboratory for Condensed Matter Physics,
and Institute of Physics, Chinese Academy of Sciences, Beijing 100190, China 
}%

\author{Bin Yan}\email{byan@lanl.gov}
\affiliation{%
 Center for Nonlinear Studies, Los Alamos National Laboratory, New Mexico, 87544
}
\affiliation{Theoretical Division, Los Alamos National Laboratory, New Mexico, 87544}

\date{\today}

\begin{abstract}
This paper presents a new way to construct single-valued many-body wavefunctions of identical particles with intermediate exchange phases between Fermi and Bose statistics. It is demonstrated that the exchange phase is not a representation character but the \textit{word metric} of the permutation group, beyond the anyon phase from the braiding group in two dimensions. By constructing this type of wavefunction from the direct product of single-particle states, it is shown that a finite \textit{capacity q} -- the maximally allowed particle occupation of each quantum state, naturally arises. 
The relation between the permutation phase and capacity is given, interpolating between fermions and bosons in the sense of both exchange phase and occupation number. This offers a quantum mechanics foundation for \textit{Gentile statistics} and new directions to explore intermediate statistics and anyons.
\begin{description}
\item[PACS numbers]
05.30.Pr, 05.30.-d, 05.70.Ce, 03.65.Vf
\end{description}
\end{abstract}

\pacs{Valid PACS appear here}
\maketitle

\section{Introduction}
Quantum statistics distinguishes its classical counterpart by virtue of the intrinsic properties of the underling quantum particles, i.e., occupations of single particle states are restricted; the many-body wavefunctions acquire phases upon particle exchanges \cite{canright90}. The latter could refer to the variable exchanges in a many-body wavefunction, or the actual adiabatic exchange of particle positions in real spaces. For fermions and bosons, these two types of exchanges are essentially the same, giving raise to geometric phases $\pi$ or $0$. Consequently, the allowed occupation numbers in a given single particle state are restricted to one or arbitrary, for fermions and bosons respectively.  

It has been realized for a long time that there exist intermediate regimes between the Fermi and Bose statistics.
Considerations of the possibility for the intermediate statistics from the view points of occupation and exchange phases are both of long history and great interest \cite{gentile40, leinaas77, wilczek82, wu84, haldane91, khare05, nayak08}. However, the relation between allowed particle occupations and the two types of exchanges phases (variable exchange or adiabatic position exchange), as well as the their implications to the statistical distributions, are far from well-understood.

For instances, Gentile statistics \cite{gentile40}, starting from allowing any no more than $q$ particles occupying the same quantum state, appeared early in 1940 but is still lack of quantum mechanics foundation; Fractional statistics (originated from studies of anyons), defined through any exchange phase $\theta$ in two dimensions \cite{wilczek82,wu84,leinaas77} attracts much attention \cite{nayak08,laughlin88,wilczek90,KITAEV20032,kitaev06} but has no explicitly deduced statistical distribution; Fractional exclusion statistics (FES) \cite{haldane91} from generalized Pauli exclusion principle can be used in anyonic system and has explicit distribution \cite{haldane91,wu94,hatsugai96}. 

Specifically, anyons are quasi-particle (or quasi-hole) excitations in two dimensional systems, e.g., the excitation of a factional quantum Hall ground state with fractional charges. The nontrivial exchange phases of anyons are generally understood as the geometric phase accumulated via adiabatic exchange between two particles \cite{canright90,wilczek82,wu84}. However, constructions of the many-anyon wavefunctions and discussions of the exchange properties are model related \cite{wu84m,pra17-many-anyon}, and the statistics is transmuted from the underlying Hamiltonian \cite{canright90,Acharya_1994,AROVAS1985117}. There were not generic many-anyon states, nor anyonic distributions were obtained. Alternatively for FES \cite{haldane91,wu94}, it is argued that in condensed matter system the one-particle Hilbert space dimension $d_\alpha$ in general depends on the occupation of states $N_\alpha$, $\Delta d_\alpha=-\sum_\beta g_{\alpha\beta}\Delta N_\beta$, as the generalized exclusion principle. The statistical parameter $g_{\alpha\beta}$ might relate to the interchange phase in some system, e.g. $g_{\alpha\beta}=\theta_{\alpha\beta}/(2\pi)$ in fractional quantum Hall effect (FQHE) \cite{haldane91} and $g_{\alpha\beta}=\delta_{\alpha\beta}\theta/\pi $ in spin chain \cite{wu94,hatsugai96} with possible verification \cite{cooper15}. There have also been a lot of discussions on their relations with Gentile statistics \cite{Han07, dai04, raj95, schoutens97, sevin07, trovato13, shen10, mirza10}. 

In this work, we tackle the problem of relating the many-body wavefunction exchange phases to the statistical distributions from a different angle. Namely, we show that there is a nature way to construct a single-valued many-body wavefunction with non-trivial phases upon variable exchanges; and intriguingly, this leads to the well-known Gentile statistics. 

The paper is structured as the following. We begin with the general requirement for phase shift of the identical particles wavefunctions in Sec. \ref{s:identical} and then construct a wavefunction with the aid of \textit{word metric} in Sec. \ref{s:many}. We prove that any interchange or permutation of the arbitrary particles yield a total phase shift on the many-body wave function. We shall note that, instead of the adiabatic phase constrained in two dimensions for anyon, the phase shift of the permutation on the many-body wavefunction can exist in any dimensions.  Applying it to the wavefunction with some particles occupying the same state in Sec. \ref{s:capacity}, we find that the interchange phase limits the maximal occupation number named \textit{capacity q} of each quantum state and that the mutual phase gives no constraint on the occupation, as a new \textit{q-exclusion principle}. Thus, the identical particles respecting permutation phase shift shall obey Gentile statistics, different from FES. Some additional discussion on the possible concern is also given.\\

\section{Exchange Phase of Identicle Particles}\label{s:identical}
Quantum mechanically, the indistinguishability of identical particles is expressed by the permutation phase shift of the many-body wavefunction:
\begin{align}
\Psi(P(q_1,q_2,\cdots, q_N))=e^{i\Theta}\Psi(q_1,q_2,\cdots, q_N),
\end{align}
where the $q_i$ denotes the quantum number of  particle-i and $P$ denotes any permutation among  those  quantum states. Given that the wavefunction is single-valued, interchanging the coordinates of two particles twice must leave the wavefunction unchanged, resulting in the only possibilities for a phase shift of $0$ or $\pi$, corresponding to bosons and fermions, respectively. 

The possibility for having arbitrary phase shift opens up once we lift the restriction of single-valueness. Multi-valued many-body functions with well-defined phase shift upon particle exchanges can exist in two dimensions \cite{wu84,leinaas77}. Indeed, this gives raise to a new type of particles, known as anyons. The existence of anyons is not merely a mathematical marvel, but has physical correspondence to quasi-particle excitations in two-dimensional systems \cite{wilczek82}. 

It is worth to emphasize that, rather than the phase shift from coordinate permutation in the wavefunction, the exchange phase of these anyon-like quasi-particles is typically referred to the geometric phase accumulated during physical exchange in the sense of adiabatic transport \cite{canright90}, without dealing with the multi-valued wavefunction.

Here, we would like to attract attention to a new possibility of having arbitrary exchange phases for a many-body wavefunction. Our approach does not assume multi-valueness of the wavefunction, nor the path dependent adiabatic exchange of particles. It is demonstrated that for single-valued wavefunctions there is indeed a consistent way to have arbitrary exchanges phases rather than $0$ and $\pi$; Namely, the permutation operator acts on the phases as well, and consequently makes the exchange phase permutation-dependent.  

Consider the composed permutation $P_2P_1$. By definition, when applied to the wavefunction $\Psi$, it generates a corresponding exchange phases $\Theta(P_2P_1)$, i.e.,
\begin{equation}
        P_2P_1\Psi = e^{i\Theta(P_2P_1)}\Psi.
\end{equation}
On the other hand, interpreting $P_2P_1$ as a sequential application of $P_1$ and $P_2$ gives an alternative expression
\begin{equation}
    P_2P_1\Psi=e^{iP_2\Theta(P_1)}P_2\Psi=e^{iP_2\Theta(P_1)}e^{i\Theta(P_2)}\Psi.
\end{equation}
This leads to a consistency condition
\begin{align}\label{id1}
P_2\Theta(P_1)+\Theta(P_2)=\Theta(P_2P_1).
\end{align}

Once equation (\ref{id1}) is satisfied, the single-valueness of the wavefunction would not be violated. \textit{This is the permutation constraint on many-body wavefunction for identical particles} we proposed. For the wavefunction as the basis of the representation of the permutation group with $e^{i\theta}$ as the character, there are only two kinds of statistics (Fermi and Bose) since the one-dimensional representation of the  permutation group can ony have  character $\pm 1$.  In the case of $P_2\Theta(P_1)\neq \Theta(P_1)$, this is \textit{not} a representation of the permutation group, beyond the fermionic and bosonic statistics.

It seems redundant \cite{leinaas77} but still straightforward and non-confusing to define the wavefunction on labeled local variables. As an analogy to Slater determinant, one can build the $N$-body wavefunction (unnormalized) as a sum over functions of permutations on variables:
 \begin{align}\label{ge-psi}
 \Psi(q_1,q_2\cdots  q_N)\equiv \sum_Pe^{i\tilde{\Theta}(P)}\Phi(P(q_1,q_2\cdots  q_N)),
\end{align}
where $\Phi$ is a $N$-variable function with $q_i$ being local variables such as coordinate numbers and $\tilde{\Theta}(P)$ is a function of the permutation $P$. Applying $P_1$ on the wavefunction, we have $P_1\Psi=\sum_Pe^{iP_1\tilde{\Theta}(P)}\Phi(P_1P)=e^{i\Theta(P_1)}\Psi$, if
\begin{align}\label{id2}
P_1\tilde{\Theta}(P)=\tilde{\Theta}(P_1P)+\Theta(P_1)
\end{align}
for any $P$. Identity (\ref{id1}) can be obtained by applying $P_2$ on both sides of equation (\ref{id2}). Without lose of generality, it is natural to let the phase of identity be zero: $\tilde{\Theta}(P_{Id})=0$, and then one can find $\tilde{\Theta}(P)=-\Theta(P)$. The price paid is that the permutation phase is label dependent. It is acceptable since this phase shift is not from any measurable physical process, but a constraint on the wavefunction. This constraint affects the many-body spectrum and statistics. \\

\section{Constructing many-body wavefunction}\label{s:many}
\begin{figure}
\includegraphics[width=\linewidth]{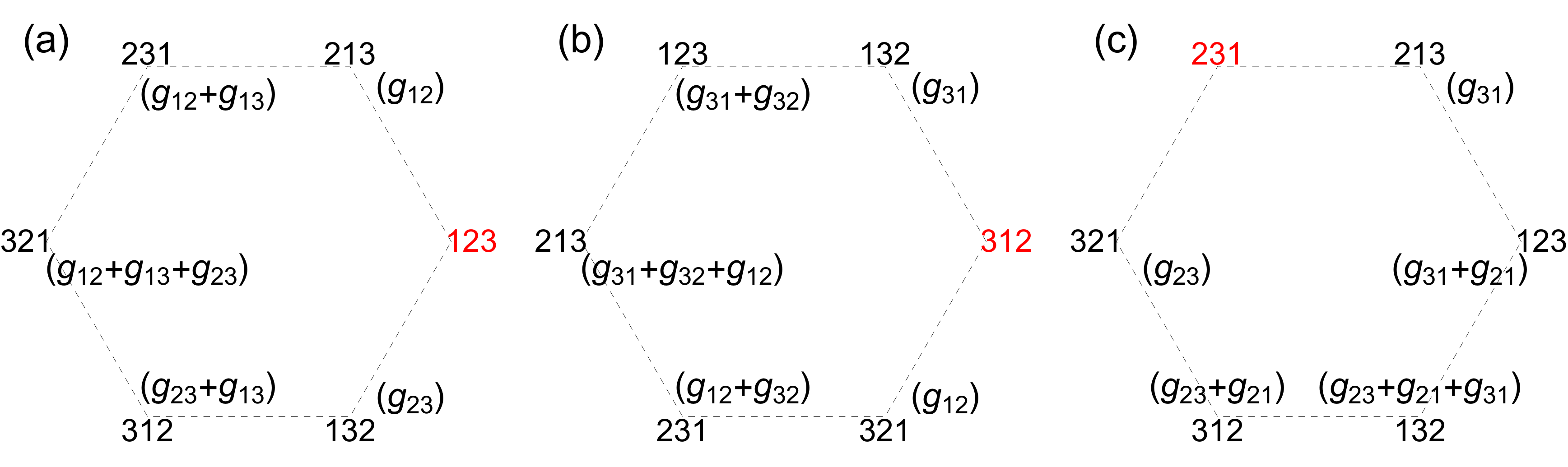}
\caption{\label{f:3p} Graphic illustration of word metric (Cayley Graph) of permutation group $S_3$. Each vertex is associated with one sequence and each edge connects two sequences different by once adjacent interchange. In (a) we enumerate all the sequences and phase factors starting from sequence $123$ (red). (b) is a cyclic permutation $1\rightarrow3\rightarrow2$ of (a). In (c) we restart from sequence $231$ (red). This is equivalent to a gauge transformation of the wavefunction.}
\end{figure}

It will be more convenient to  represent each permutation $P$ with a (labeled particles) sequence, $a_1a_2 \cdots a_N$. The identity operator is the natural ordering sequence: $P_{Id}=12\cdots N$. After applying $P$, the  sequence becomes  $a_1a_2\cdots a_N$. Define the pair parameter $g_{mn}=Sgn(n-m)$ and
\begin{align}\label{length}
l(P)=\sum_{(mn)} g_{mn},
\end{align}
here $(mn)$ means inversion pair: $m$ appears on the right of $n$ for $m<n$ in sequence $P$. From sequence $P_{Id}$ to $P$, particle-$m$ and -$n$ must be interchanged, so the number of inversion pairs equals to the number of least adjacent interchange \cite{BH01} (least \textit{braiding} in two dimension) , called the \textit{length} of $P$ in \textit{word metric} \cite{billey07}. Any permutation can be decomposed as least adjacent interchange, so the phase of $P$ could be interpreted as $\Theta(P)=\theta l(P)$ with $\theta$ the characteristic interchange phase of this kind of identical particles,   alike to the anyon in the sense of any interchange phase. Thus the wavefunction is 
\begin{align}\label{psi}
\Psi\equiv\sum_Pe^{-i\theta l(P)}\Phi(P).
\end{align}
 When $\theta=0/\pi$, it recovers the bosonic/fermionic wavefunction. Explicitly, the wavefunction for two-particle is
\begin{equation}\label{2}
\Psi= \Phi(1,2)+e^{-i\theta g_{12}}\Phi(2,1).
\end{equation}
Interchange particle -$1$ and -$2$ denoting as $T^{12}$, 
\begin{align}
T^{12}\Psi=\Phi(2,1)+e^{-i\theta g_{21}}\Phi(1,2)=e^{-i\theta g_{21}}\Psi,
\end{align}
and certainly repeated interchange
\begin{align}
T^{12}(T^{12}\Psi)=e^{-i\theta g_{12}}e^{-i\theta g_{21}}\Psi=\Psi
\end{align} 
leaves the wavefunction unchanged, respecting the single-valueness of the wavefunction. As mentioned before, the essence is that $T^{ij}$ also changes $g_{nm}$. A three-particle example can be visualized in Fig.\ref{f:3p}. Diagram (a) and (b) enumerates all the sequences and the corresponding phase terms for $\Psi$ and $T^{132}\Psi$, with cyclic permutation $T^{132}$ meaning $1\rightarrow3\rightarrow2$. Comparing the phase difference of the same sequences before and after the permutation, the wavefunction feels a phase shift:  
\begin{align}
T^{132}\Psi=e^{i\theta(g_{13}+g_{23})}\Psi=e^{i\theta l(T^{132})}.
\end{align}
The general phase shift $P\Psi=e^{i\theta l(P)}\Psi$ is proven in Appendix (\ref{s:app1}).

Not only the phase dependent on the artificial labels, it is also sensitive to the selection of zero phase sequence. Different selection means to re-define a new origin rather than the identical operator in word metric and $l(P)$ would change correspondingly. Yet, this does \textit{not} hurt the metric, indistinguishability and statistics since re-selection is equivalent to a permutation of particles and the wavefunction is shifted by a phase as seen in the example of three particles case in Fig. \ref{f:3p}(c). Starting from sequence $231$ and defining $l(P)$ as the sum of the inverted pair parameters referring to sequence $231$, the phase shift is $\Psi^\prime =e^{i\theta l(231)}\Psi$.

Typically, $l(P_1P_2)\neq l(P_1)+l(P_2)$ and $e^{-i\theta l(P)}$ is \textit{not} a representation character of the permutation group, unless $e^{-i\theta}=\pm1$. Moreover, $l(P_1P_2)\neq l(P_2 P_1)$ in general, so the phase shift is indeed non-Abelian. We shall clarify that this is due to the property of word metric of non-Abelian permutation group, rather than the path-dependent braiding like non-Abelian anyons \cite{MOORE91-nonabelians} or charges \cite{Wu-sci19}.

In the coordinate space $X$, coordinates can be used as labels: $g(x,y)=Sgn(y-x)$. The sign function could be generalized in high dimensions, since the Cartesian product of \textit{totally ordered} set is still totally ordered. For example, the $arg(z)$ function in two dimension \cite{wu84m} or $Sgn(\sum_i(Sgn(y_i-x_i)))$ in any dimension. The ``\textit{q}-symmetrize" \cite{mitra95,goldin04} wavefunction for two particles $\phi(x,y)+q(x,y)\phi(y,x)$ is the same to ours with $q(x,y)=e^{i\theta g(x,y)}$. Yet, without the help of word metric, the many-body permutation properties and statistics can not be explored.

\section{Gentile Statistics}\label{s:capacity}
To explore the occupation of each quantum state, the wavefunction should be constructed from the direct product of single particle states: $\Phi(P)=\phi_a^1\phi_b^2\cdots\phi_c^N$ where $\phi_a^i$ is the wavefunction of particle-$i$ at state-$a$, which is impossible for the path-dependent braiding anyon exchange phase \cite{wilczek90,pra17-many-anyon}. Consider two and three particles occupying the same state (ignoring the particle labels for them are at the same state), the wavefunctions become 
\begin{align}
\Psi_{2a}=&(1+e^{-i\theta})\phi_a\phi_a,\label{2same}\\
\Psi_{3a}=&(1+e^{-i\theta})(1+e^{-i\theta}+e^{-2i\theta})\phi_a\phi_a\phi_a.
\end{align}
By induction (see Appendix. \ref{s:app2} ) the wavefunction of $N_a$ particles at the same state-$a$ is
\begin{align}\label{manysame}
\Psi_{N_a}=&\prod_{n=0}^{N_a-1}(\sum_{k=0}^ne^{-i*k\theta})\times\phi_a...\phi_a\nonumber\\
=& \prod_{n=1}^{N_a}(\frac{1-e^{-i*n\theta}}{1-e^{-i\theta}})\times\phi_a...\phi_a.
\end{align} 
Indeed, this is the \textit{Q-factorial} form for the number operator in quantum group \cite{macfarlane89} with $Q=e^{-i\theta}$. The Q-factorial of $n$ is defined as $n!_Q\equiv \prod_{i=0}^{n-1}(\sum_{k=0}^i Q^k)$. Analogy to the exclusion of fermion, where the wavefunction of fermions occupying the same state vanishes, i.e., $1+e^{-i\pi}=0$, the many-particle wavefunction (\ref{manysame}) vanishes when $N_a\geqslant q+1$, providing $(e^{-i\theta})^{1+q}=1$. As a consequence, the maximal occupation number named \textit{capacity} of state-$a$ is $q$. 

In the (real) Q-analogy quantum group approach \cite{macfarlane89,fivel90,greenberg91,Chung2017} ($aa^\dagger +Qa^\dagger a=Q^N$ with $-1<Q<1$), it is noticed that the number operator is positive definite unless $Q$ is the root of unity (impossible in their case). If we were allowed to extend the real $Q$ to the interchange phase $e^{-i\theta}$, the capacity $q$ of quantum state should be found.

In the case that $\theta/\pi$ is a simple fraction $r/p$, where the thermodynamic limit can be achieved via a sequence of systems with different particle numbers \cite{haldane91,wilczek82}, the phase-capacity relation is explicit:
  \begin{align}\label{capacity}
    q=\left\{
                \begin{array}{ll}
                  2p-1,\ \ r\  is\  odd\\
                  p-1,\ \ \ r\  is \ even.\\
                \end{array}
              \right.
  \end{align}
This is the \textit{q-exclusion principle} we obtained from the permutation constraint on the wavefunction. The capacity is degenerate due to the $r \pm 2k$ freedom on the numerator. Alike to the accuracy of plateau resistance in FQHE \cite{tsui82,stern08}, $\theta/\pi$ shall be exactly a fraction. In Table \ref{t:p-q} we list the fractions for several small capacity numbers and intriguingly they are the prominent filling fractions in FQHE \cite{jain-prl89,jain07composite}. Taken the smallest positive fractions for each capacity numbers, we also note that the phase-capacity relation is monotone decreasing, smoothly interpolating between fermion and boson in the sense of both exchange phase and occupation numbers.  Besides, to obtain this exclusion in real space, namely the possibilities of particles at different states staying the same position, we shall let $g(x,x)=1$.

\begin{table}[t]
\caption{\label{t:p-q}Some examples of phase-capacity relation from equation (\ref{capacity}). The capacities are degenerate some fractions due to the $r\pm 2k$ on the numerator and we take the smallest positive fractions. }
\begin{ruledtabular}
\begin{tabular}{c|cccccccc}
$\theta/\pi$&1&2/3&1/2&2/5&1/3&2/7&$\cdots$&0\\
$q$&1&2&3&4&5&6&$\cdots$&$\infty$ \\
\end{tabular}
\end{ruledtabular}
\end{table}

Many-particle wavefunction with mutual exchange phases $\theta_{\alpha\beta} $ among mixed species could be similarly achieved by replacing $\theta l(P)$ in equation (\ref{psi}) with $\Theta(P)=\sum \theta_{\alpha_m\alpha_n}g_{mn} $, where $\alpha_m$ is the specie particle-$m$ belonging to. For $N_a$ particles at state-$a$ of specie $\alpha$ and $N_b$ particles at state-$b$ of specie $\beta$,
\begin{align}\label{mut}
\Psi_{N_a^\alpha N_b^\beta}=&\prod_{n=0}^{N_a-1}(\sum_{k=0}^ne^{-i*k\theta_{\alpha\alpha}})\prod_{n=0}^{N_b-1}(\sum_{k=0}^ne^{-i*k\theta_{\beta\beta}})\nonumber\\&
\times\sum e^{-i\Theta(P)}P\phi_\alpha^{N_a}\phi_\beta^{N_b}.
\end{align}
In the last sum of permutation terms, there are $(N_a+N_b)!/(N_a!N_b!)$ distinct permutations and all the mutual exchange phases $\theta_{\alpha\beta} $ appear in these terms. They are mutual orthogonal and the superposition will never cancel. Thus the wavefunction can only vanish due to the front Q-factorial terms from the permutation phase of the same state in the first line. Therefore, neither the occupation numbers of the same species different states ($\alpha=\beta,\ a\neq b$) nor of the different species ($\alpha\neq \beta$) mutually affect.  Only the interchange phase of the same specie at the same state constrains the capacity of each single state, quite different from the mutual statistics parameter $g_{\alpha\beta}$ from FES \cite{haldane91,wu94}.\\

Finite capacity of each quantum state means Gentile statistics \cite{gentile40}.   The thermodynamics of Gentile statistics such as the heat capacity, condensation, equation of state and relation with FES are well studied \cite{khare05,dai04,lavagno00,shen10}. At zero temperature, the occupation number $\langle n_i \rangle =q$ for those states below the Fermi-like surface, similar to the FES without mutual statistics $g_{\alpha\beta}=\alpha\delta_{\alpha\beta} $ \cite{wu94}. Plausibly, one may consider that $N_j$ particles occupy at least $\lceil N_j/q\rceil$ (ceiling function) states among $G_j$ states. Thereafter the effective Fock space dimension is $d_B=G_j+N_j-\lceil N_j/q\rceil\approx G_j+(N_j-1)(1-1/q)$ and the number of ways $W^\prime_j=C^{N_j}_{d_B}$ is the same to Wu's counting \cite{wu94} for FES, providing $q=1/\alpha$. Here the Binomial constant $C^k_G\equiv G!/(k!(G-k)!)$. Indeed, the exact statistical weight of assigning $N_j$ particles into $G_j$ states with capacity $q$ (\textit{combination with limited repetition} \cite{lavenda91}): 
\begin{equation}\label{we}
W_j=\sum_{k=0}^{k=\lfloor N_j/(q+1)\rfloor}(-1)^kC^k_{G_j}*C^{G_j-1}_{G_j-1+N_j-k(q+1)}.
\end{equation}
The most probable occupation  number $\bar{n}_j$ is determinant from identity \cite{wu94} $\delta_{n_j}\log W_j|_{\bar{n}_j}+\log z_j=0$ with $z_i\equiv e^{-\beta(\epsilon_i-\mu)}$ being the fugacity coefficient and $\beta,\epsilon_i,\mu$ being the inverse temperature, energy level and chemical potential, respectively. It is well known \cite{dai04,shen10} that the Gentile statistics and the FES do not produce the same distribution.
 
Moreover, the relation between the statistics parameters and the occupation number is different for the Gentile statistics and FES. In the generalized ideal gas of FES ,the statistics parameter is mapped to the phase by $\alpha=\theta/\pi=r/p$ \cite{wu94}, so the occupation $\bar{n}^\prime_{\epsilon<\varepsilon_F}$ is $p/r$, not necessary an integer. This is different from the capacity  we found for the identical particles. For instance, when $\theta=2\pi/3$, the state capacity is $q=2$ while by FES, $\bar{n}^\prime_{\epsilon<\varepsilon_F}=3/2$ and for $\theta=\pi/2$, we get $q=3$ while $\bar{n}^\prime_{\epsilon<\varepsilon_F}=2$.

\section{Discussion}
It appears confusing to talk about many particles at the same state respecting $P\Psi=e^{i\theta l(P)}\Psi$, since people might also expect $P\Psi=\Psi$ for permutation of particles at the same state seems  to change nothing. Thereafter $\Psi=e^{i\theta l(P)}\Psi$ and the wavefunction seems to vanish for $\theta \neq 0$, resulting in \textit{hard-core condition}. In some proposed models, such as the magnon excitation in Heisenberg spin chain via Bethe \textit{ansatz} \cite{hatsugai96,karbach00}, the charge-flux model \cite{wu84m} and the spin-anyon mapping from Jordan-Wigner transformation \cite{fradkin89,batista01}, the hard-core condition is always found or preinstalled, resulting in the conventional Fermi distribution \cite{hatsugai96}. Indeed in the coordinate space $X^N/S_N$, the wavefunction is multi-valued \cite{wu84m,leinaas77} in the sense of permutation phase shift. The hard-core condition is a too strong constraint for identical particles \cite{leinaas77} and the coinciding points (some $x_i$'s are equal) in our wavefunction are still singular since only finite of them could be equal. Besides,  phase shifted defined in our wavefunction is a result of the \textit{ Gedanken} permutation of particles, rather  than the phase evolution of any dynamic process in low dimensions. Thus, this phase shift is not limited in spacial two dimensions, neither measurable during any adiabatic process nor interference of  different paths.   The influence of this phase buries into the q-exclusion principle and thus can be detected in any non-fermion/boson distribution related experiments \cite{Mohammadzadeh_2017,Yurovsky-prl17,Andrade-prl10,marinho2019intermediate}.

In summary, we proposed a consistency condition (\ref{id1}) for the single-valueness of the many-body wavefunction. Constructed similar to the Slater determinant, with the aid of word metric of the permutation group, our wavefunction guarantees the complete permutation properties $P\Psi=e^{i\theta l(P)}\Psi$ and gives a straightforward phase-capacity relation as concluded in equation (\ref{capacity}).  This could be the quantum mechanics foundation for Gentile statistics and generalization of quantum statistics for identical particles. New many-body features may be found under our method of wavefunction.\\

\begin{acknowledgments}
This work was initiated when both authors were graduate student at Purdue University. QZ wishes to thank Professor Yong-Shi Wu for discussion and encouragement of this work. Now QZ is supported from the International Young Scientist Fellowship of Institute of Physics CAS (Grant No. 2017002) and the Postdoctoral International Program (Y8BK131T61) from China Postdoctoral Science Foundation. BY has been supported by the U.S. Department of Energy, Office of Science, Basic Energy Sciences, Materials Sciences and Engineering Division, Condensed Matter Theory Program, and in part by the Center for Nonlinear Studies at LANL.
\end{acknowledgments}

\appendix
\section{Interchange properties}\label{s:app1}
To explore the permutation phase of the wavefunction, we decompose permutation into least adjacent interchange $\sigma_i$'s called \textit{reduced word} \cite{billey07}, where $\sigma_i$ denotes the interchange of the digits at position-$i$ and $i+1$ \cite{wu84}. Although the decomposition is not unique, the inversion pairs are determinate. For example, $P$ of sequence $2413$ can be decomposed as $P=\sigma_2\sigma_3\sigma_1=\sigma_2\sigma_1\sigma_3$, while the inversion pairs are $(12),(14),(34)$ and $l(2413)=g_{12}+g_{14}+g_{34}$. 

An example of permutation phase shift for three-body wavefunction is illustrated in Fig. \ref{f:3p}. To prove $P\Psi=e^{i\theta l(P)}\Psi$ is to verify condition (\ref{id2}), or $l(PP_1)-P(P_1)=l(P)$ for any $P,P_1$. Since every permutation could be decomposed into independent (commutative) cyclic permutations, the general justification of any permutation on wavefunction could be concluded from the following two cases.

Firstly, consider the 2-circle permutation $T^{ij}$. The following two sequences($i<j$)
\begin{align*}
P_1=\underbrace{a_1a_2...a_{k-1}}_{s_1}\ i\ \underbrace{a_{k+1}a_{k+2}..a_{l-1}}_{s_2}\ j\ \underbrace{a_{l+1}a_{l+2}..a_N}_{s_3},\\
P_2= \underbrace{a_1a_2...a_{k-1}}_{s_1}\ j\ \underbrace{a_{k+1}a_{k+2}..a_{l-1}}_{s_2}\ i\ \underbrace{a_{l+1}a_{l+2}..a_N}_{s_3},
\end{align*} 
 with length
\begin{align*}
l(P_1)=&\sum_{m_1>j}g_{jm_1}+\sum_{m_1>i}g_{im_1}+\sum_{m_2>j}g_{jm_2}+\sum_{m_2<i}g_{m_2i}\nonumber\\
&+\sum_{m_3<j}g_{m_3j}+\sum_{m_3<i}g_{m_3i}+L,\\
l(P_2)=&\sum_{m_1>i}g_{im_1}+\sum_{m_1>j}g_{jm_1}+\sum_{m_2>i}g_{im_2}+\sum_{m_2<j}g_{m_2j}\nonumber\\
&+\sum_{m_3<i}g_{m_3i}+\sum_{m_3<j}g_{m_3j}+g_{ij}+L,
\end{align*}
satisfy $T^{ij}P_1=P_2$ and $T^{ij}P_2=P_1$, where $m_i$($i=1,2,3$) are the digits within segment $s_i$ and $L$ is the sum of terms independent from $ij$. The phase shift for the two sequences during mapping shall be equal:
\begin{align}
l_{ij}\equiv T^{ij}l(P_2)-l(P_1)=T^{ij}l(P_1)-l(P_2).
\end{align}
The total wavefunction feels a phase shift when the above identity is independent of the position of particles-$i,j$ and the ordering of other particles. The only solution for the above identity is $g_{mn}=-g_{nm}$ and the overall phase shift is $e^{-i\theta l_{ij}}$ with
\begin{align}
l_{ij}=\sum_{m>i}^{m<j}(g_{mi}+g_{jm})+g_{ji}.
\end{align}
$T^{ij}$ itself is a 2-\textit{circle} permutation as an element of the permutation group and it could be represented as sequence
\begin{align}
12\ast\ast(i-1)j(i+1)\ast\ast(j-1)i(j+1)\ast\ast N.
\end{align}
with the starts parts ``$\ast\ast$" being the natural increasing ordering. The length of $T^{ij}$ is
\begin{align}
l(T^{ij})=&\sum_{m>i}^{m<j}(g_{im}+g_{mj})+g_{ij}\\
=&-l_{ij}.
\end{align}
Thus as cited in the main text,
\begin{align}
 T^{ij}\Psi= e^{i\theta l(T^{ij})}\Psi.
\end{align}

Secondly, in a more complicated case the permutation could be decomposed as two independent cyclic permutation, $P=T^{ijk}T^{mn}$ with, say, $i<k<m<j<n$. Applying this $P$ on $\Psi$, the sequence
\begin{align*}
S_1=\cdots i\cdots m\cdots j\cdots n\cdots k\cdots
\end{align*}
maps to 
\begin{align*}
S_2=\cdots j\cdots n\cdots k\cdots m\cdots i\cdots,
\end{align*}
but not inversely. The dots parts ``$\cdots$" for the  sequences $S_1$ and $S_2$ are the same. Now the phase shift from sequence $S_1$ to $S_2$ is $e^{-i\theta l_{ijk,mn}}$ with
\begin{align}
l_{ijk,mn}\equiv T^{ijk}T^{mn}l(S_1)-l(S_2).
\end{align}
Once $g_{nm}$ is antisymmetric, the phase shift is the same for all the ordered  sequences in $\Psi$ and
\begin{align}
l_{ijk,mn}=&\sum_{a>i}^{a<j}g_{ja}+\sum_{a>k}^{a<j}g_{ak}+\sum_{a>i}^{a<k}g_{ai}+\sum_{a>m}^{a<n}g_{na}+\sum_{a>m}^{a<n}g_{am}\nonumber\\
&+(g_{ij}+g_{kj}+g_{mj}+g_{kn}+g_{mn}).
\end{align}
The first five terms sum  over all the numbers except $i,j,k,m,n$. The first three of them are the inversion pair parameters for $T^{ijk}$ and the following two are for $T^{mn}$. The last bracket term is the inversion pairs parameters among $i,j,k,m,n$ themselves. Likewise, permutation $T^{ijk}T^{mn}$ is able to represented as sequence
\begin{align}
12\ast\ast j\ast\ast i\ast\ast n\ast\ast k\ast\ast m\ast\ast N,
\end{align}
Then by definition (\ref{length}), we can also find $l(T^{ijk}T^{mn})=-l_{ijk,mn}$. Longer circle and more complicated decomposition would not make things different. It thus is not hard to generalize to any permutation $P$,
\begin{align}\label{permute}
P\Psi=e^{i\theta l(P)}\Psi.
\end{align}

As to mutual statistics, replace $\theta l(P)$ with $\Theta(P)=\sum \theta_{\alpha_i\alpha_j} $ and let $\theta_{\alpha_i\alpha_j}$ be antisymmetric, here $\alpha_i$ is the species particle-$i$ belonging to, the general permutation property (\ref{permute}) becomes
\begin{align}
P\Psi=e^{i\Theta(P)}\Psi,
\end{align}
in the exactly same spirit.\\
 
\section{Occupying the same states}\label{s:app2}
For $N$-particle sequences, denote $I_N(l)$ as the number of permutation sequences with $l(\leq C^2_N)$ inversion pairs. Clearly, $I_N(0)=1$ and define the generating function $\Psi_N(x)$ of $I_N(l)$:
\begin{align}
\Psi_N(x)\equiv\sum_{l=0}^{C^2_N}I_N(l)x^l.
\end{align}
It satisfies the recurrence relation \cite{BH01}:
\begin{align}
\Psi_N(x)=(1+x+x^2+\cdots+x^{N-1})\Psi_{N-1}(x)\label{rec}.
\end{align}
For $N_a$-particle at the same state, wavefunction (\ref{psi})
\begin{align}
\Psi_N&=\sum_Pe^{-i\theta l(P)}P\phi_a...\phi_a \\
&=\sum_{l=0}^{C^2_N}I_N(l)(e^{-i\theta })^l\phi_a...\phi_a
\end{align}
is indeed the generating function of $I_N(l)$ with argument $e^{-i\theta}$. $\Psi_1=1$ and by recurrence relation (\ref{rec}),
\begin{align}
\Psi_{N_a}=&\prod_{n=0}^{N_a-1}(\sum_{k=0}^ne^{-i*k\theta})\times\phi_a...\phi_a\label{Q-form}\\
=& \prod_{n=1}^{N_a}(\frac{1-e^{-i*n\theta}}{1-e^{-i\theta}})\times\phi_a...\phi_a.
\end{align} 

For the wavefunction of many particles involving different states or different species with mutual interchange phases, the wavefunction can be checked out in the same way. Preset the order $\Phi_0=\phi^1_a\cdots\phi^{N_a}_a\phi^{N_a+1}_b\cdots\phi^{N_a+N_b}_b$ with zero phase and permute on the same state first, the Q-factorial like phase terms for the same state appear. Then we permute on the different states to get the mixed distinct terms and the wavefunction with different states or species shall take the form:
\begin{align}
\Psi_{N_a^\alpha N_b^\beta}=&\prod_{n=0}^{N_a-1}(\sum_{k=0}^ne^{-i*k\theta_{\alpha\alpha}})\prod_{n=0}^{N_b-1}(\sum_{k=0}^ne^{-i*k\theta_{\beta\beta}})\nonumber\\&
\times\sum e^{-i\Theta(P)}P\phi_\alpha^{N_a}\phi_\beta^{N_b}.
\end{align}


%

\end{document}